# Surface and Mechanical studies of Bismaleimide coatings

A.S.Bhattacharyya[1]; D.Paul [2]; P.P. Dutta [2]; G.Bhattacharjee[2]
[1]*Centre for Nanotechnology, Central University of Jharkhand, Brambe, Ranchi 835205, India*
[2]*Department of Mechanical Engineering, Sikkim Manipal Institute of Technology Sikkim 737136, India*

*Bismaleimide (BMI) resins are a new breed of thermosetting resins used mainly for high temperature applications and have major usage in aerospace. BMI polymer coatings were deposited on aluminum and mild steel substrates. The effect of corrosion on mild steel and aluminum by Ringers Solution and there protection using BMI coatings were observed. X-ray diffraction studies showed crystalline nature of the BMI coatings. Surface contact angle measurements were carried out using goniometer.*

e-mail: arnab.bhattacharya@cuj.ac.in; 2006asb@gmail.com

In our previous publication we have discussed about structural configuration of Bismaleimide (BMI) resins and formation of metallized trilayers of Bismaleimide (BMI) resins and its morphology [1]. BMI are thermosetting polymers having properties of dimensional stability, low shrinkage, chemical resistance, fire resistance, good mechanical properties and high resistance against various solvents, acids, and water [2, 3]. Details regarding BMI can be found in literature. Polymers coatings on the other hand are applied on steel, woods and other metal parts to improve its corrosion resistance and other environmental attacks. They are mainly used in the automobile industry. These coatings are very helpful for protection for steel pipelines used in gas and oil supply and emit zero or near zero volatile organic compounds. There are lots of US patents on the deposition of polymer powder coatings on metals [4, 5]. BMI is commercially available as Homide 250[6]. BMI coating has also been used for the corrosion protection [7].

## MATERIALS & METHODS

**Preparation of substrates**
Samples of aluminum and mild steel were made from aluminum and mild steel sheets of thickness 1mm and 2 mm respectively. These samples were first rubbed and cleaned with emery paper to remove a part of atmospheric rust and then they were polished, hammered to flatten and cut into small pieces as shown in figure 1 & 2 which were used as substrates for the deposition of BMI coatings.

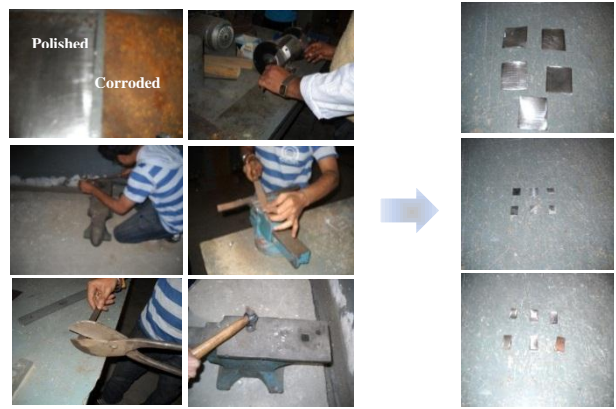

**Fig 1** Mild steel sheets were polished, hammered and cut into small pieces

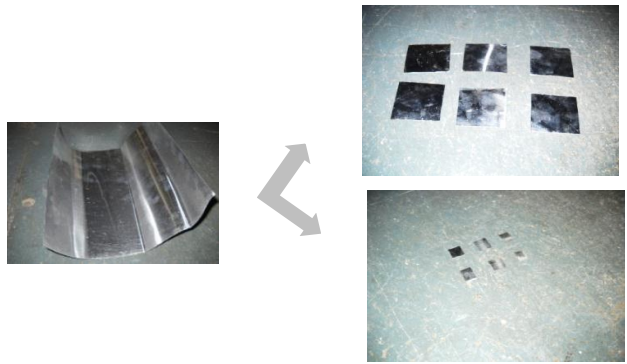

**Fig 2** Aluminum substrates were cut into small pieces

## Vacuum deposition of aluminum coatings

Details regarding formation of BMI coatings are given in our earlier publication [1]. Aluminum coatings were deposited on BMI coated mild steel and aluminum substrates. The coating unit was supplied by Vacuum Equipment Company, India. The process consists of resistive heating of a small Al strip which gets evaporated and gets deposited on the substrate kept normally above the heating coil (Fig 3). The chamber is evacuated to a vacuum of 0.05 mbar. The vacuum is required to avoid collision of the metal atoms with the air molecules during their path toward the substrate for deposition and also protects the substrate from getting coated with contaminants [8].

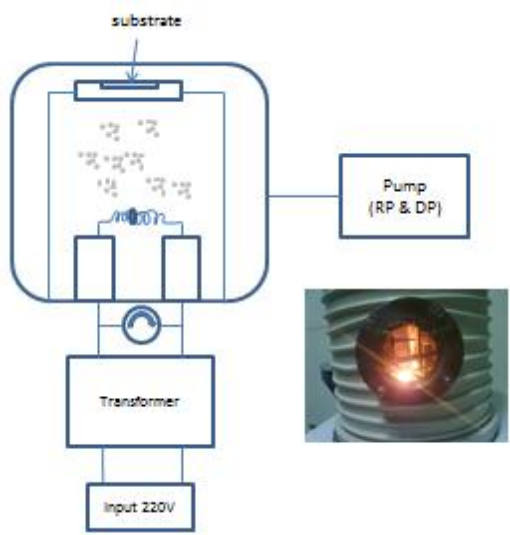

**Fig 3**: Vacuum thermal coating using resistive heating

## Corrosion studies

Corrosion of metals refers to electrochemical oxidation. Oxygen is the most common oxidizing agent which leads to the formation of iron oxide on the surface of iron, commonly known as rust. Metals exposed to saline environments also undergo corrosion and is a matter of concern for industries. Polymers on the other hand are far more resistant to corrosion than metals. Therefore polymer coatings on metal can serve the purpose of protection of the underlying metal parts. Corrosion studies were performed by treating the bare metal substrates as well as BMI coated metals in Ringers solution (Fig 4). The Ringers solution was prepared by mixing 8.6 g NaCl, 0.3 g KCl and 0.33 g $CaCl_2$ in boiled distilled water [9]. The samples were treated with acetone for about 48 hrs to remove the rust that has penetrated inside the surface. Acetone treated samples are then allowed to react with Ringer's solution for different time periods. The condition of these samples are analyzed and compared on different time periods. Both bare and BMI coated Al and MS sheets were immersed in the Ringers solution for 15 days and periodic observations were made.

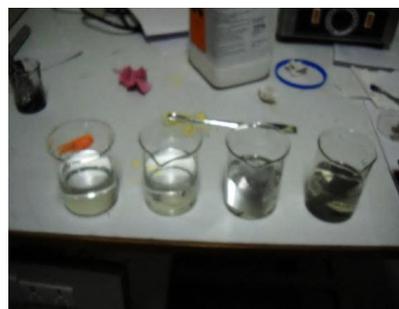

**Figure 4:** Samples kept in Ringers solution

## Surface profiles

Talysurf series 2 by Taylor Hobson at National Metallurgical Laboratory (CSIR), Jamshedpur was used for measurement of thickness and roughness of the coating. A position sensitive and a motion sensitive gauge transducer consisting of piezoelectric sensor converted the tip movement into electrical signal which got amplified to produce corresponding vertical and horizontal magnification resulting in an surface topography of the coating [10].

The "surface" is defined as the boundary between material and air. The most commonly used parameter to define surface roughness is average roughness ($R_a$).

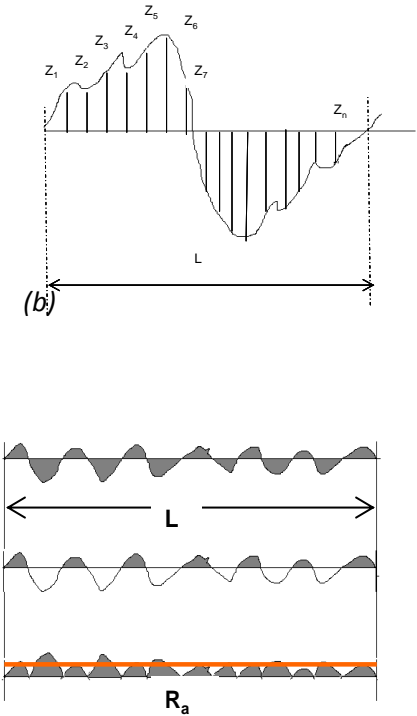

**Fig 5**: Measurement of average roughness [10, 11]

$R_a$ is defined mathematically as the absolute (or modulus of the) departure of the profile from the reference line to the sampling length as shown in *Fig 5(a, b)* and given mathematically as

$$R_a = \frac{|z_1| + |z_2| + \cdots\cdots + |z_n|}{n} \quad [10].$$

Thickness of the coating is another parameter which is related to the initial deposition conditions and the nature of the substrate. The thickness measurement was performed from a step obtained using surface profilometer at the interface between the coated and uncoated substrate. The step height was the thickness of the coating.

**Tensile tests**
Universal testing machine is a type of machine which measures the mechanical properties of a material. Mostly it measures Young's modulus, yield strength, ultimate tensile strength, shear strength. These properties are measured by tensile test and shear strength. In the tensile test a specially prepared specimen is subjected to continuously increasing uni-axial tensile force. The elongation of the material is measured simultaneously to obtain the stress-strain curve for the specimen. The stress can be either tensile or compressive. We have used tensile tests for aluminum and mild steel samples.

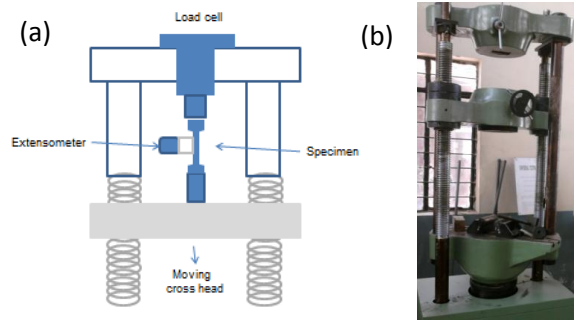

**Fig 6:** Schematic representation of universal tensile machine (UTM)

There are three principle ways in which a load may be applied for mechanical testing: Tension, compression and shear. The UTM machine in tension mode is designed to elongate the specimen at a constant rate and continuously and simultaneously measure the instantaneous applied load (with a load cell) and the resulting elongations (extensometer).

A rectangle bar of aluminum and mild steel was first brought to a specified shape according to ASTM standard [12]. The bar should be broad at the ends and narrow at the middle which is also known as dog- bone shape (Fig). During testing, the deformation should confine in the narrow central region called the gauge length, which has a uniform cross section along its length (fig c). This specimen was then subjected to tensile test to obtain the corresponding stress-strain curve. The machine is hydraulically operated by a pump.

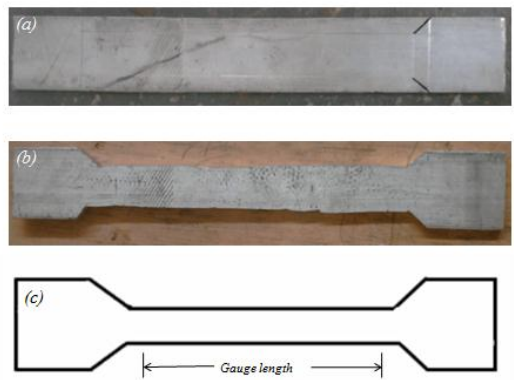

**Fig 7**: ASTM standard tensile testing specimen

Surface morphology was studied by Atomic Force Microscopy (AFM) by Nanosurf 2.and Bruker AXS was used at IACS Kolkata was used for the XRD studies where monochromatic Cu-K$_\alpha$ radiation having a wavelength of 1.54Å was used. Vickers microhardness tester supplied by Leica (VMHT Auto), Germany at National Metallurgical Laboratory (CSIR), Jamshedpur was used to measure the hardness and goniometer was used for contact angle measurements

## RESULTS & DISCUSIONS

**Corrosion studies**
The Homide 250 BMI coatings on aluminum were hard and had good adherence to the aluminum substrate details of which are published in ref [1]. The effect of corrosion by putting the mild steel substrates into Ringers solution is shown in Fig 3.1. It can be clearly observed that the effect of corrosion in a saline environment is very severe.

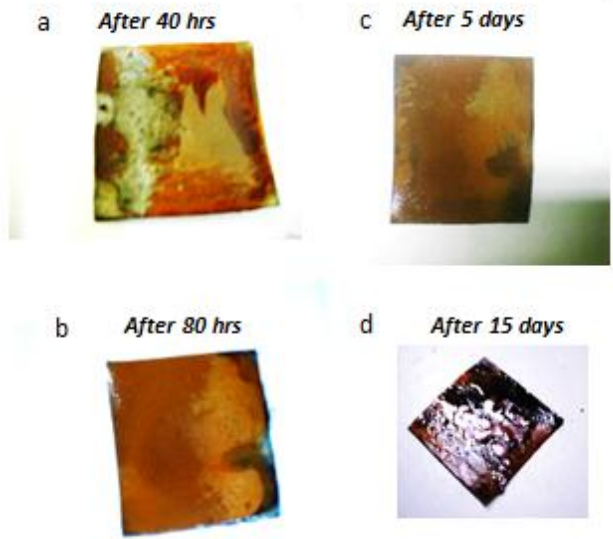

**Fig 8**: The effect of corrosion on mild steel immersed in ringers solution for (a) 40 hrs
(b) 80 hrs (c) 5 days and (d) 15 days

Aluminum is corrosion resistant compared to MS due to the formation of protective $Al_2O_3$ layer on the substrate which prevents further oxidation. However, prolonged exposure to saline environment (Ringers solution) even causes corrosion in aluminum as shown in Fig 8 (a). BMI coated aluminum however showed protection from corrosion even after prolonged exposure to saline environments as shown in fig 8(b).

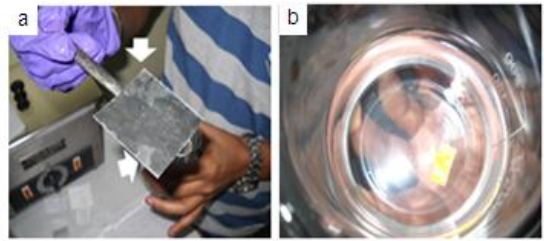

**Fig 9**: (a) Corrosion on aluminum substrates and (b) no signs of corrosion on BMI coated substrates

**Vickers microhardness**
Vickers microhardness test of the aluminum and mild steel substrate at different loads are plotted and shown in Fig 10 (a, b). A decrease in hardness with increase in load is observed. This phenomenon is called indentation size effect. In case of mild steel substrate however the hardness initially shows an increase in value with increase in load and later on shows a decrease. The initial increase is due to work hardening a phenomenon which occurs due to entanglement of dislocations. An optical micrograph of the indentation made on Al substrate at 25gf is shown in fig 11.

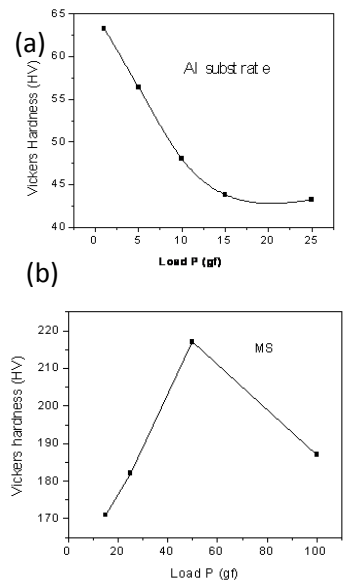

**Fig 10**: Vickers microhardness with different loads on (a) aluminum and (b) Mild steel substrates

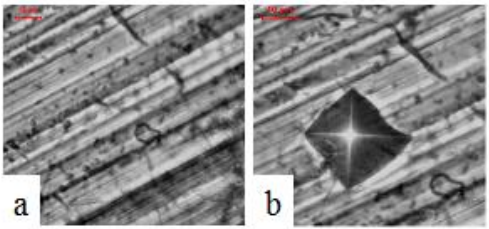

**Fig 11**: Optical micrograph of (a) surface of aluminum substrate and (b) indentation made at a load of 25 gf

Microhardness test on aluminum substrate and an Al-BMI-MS trilayer at 25 gf were done. The Al and MS substrate showed a hardness of 43.2 HV and 182 HV respectively. The trilayer showed a substantial increase in hardness (780 HV) as shown in fig 12 and reported in [1].

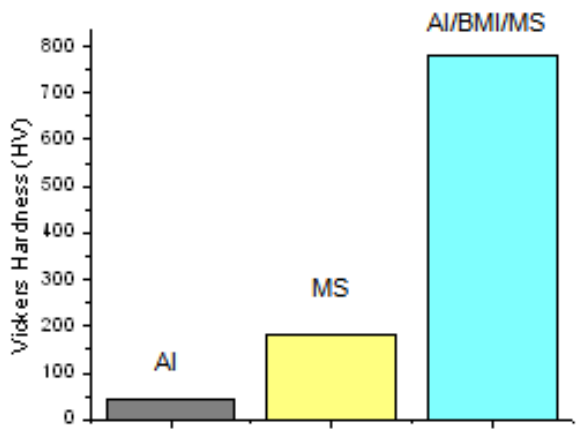

**Fig 12**: Vickers microhardness performed on aluminum and Al-BMI-MS at 25 gf[1]

### Surface profilometer

The average roughness of the corroded mild steel substrates was found to be much higher than plane mild steel due to formation of rust (Figure 13). The average roughness Ra was 717nm for plane mild steel substrate and was 1.74 µm for corroded mild steel substrates. The thickness of the BMI coatings was in micrometer range and varied from 17µm to 100 µm depending upon the deposition time. Surface profiles indicating BMI thickness is shown in fig.14. Coatings of higher thickness usually get peeled off from the substrates due to stress arising due to mismatch of coefficient of thermal expansion (CTE). Bare Al substrate was found to have a roughness of about 100nm. The value increase to 800nm in case of BMI deposited on Al substrates. This increase in roughness will result into better adhesive properties of the BMI surface.

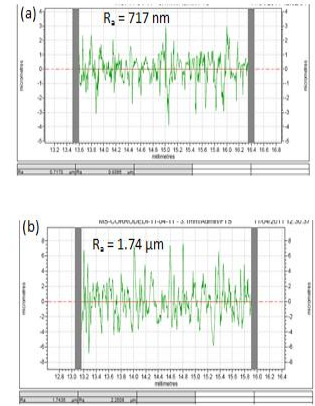

**Fig 13:** Roughness profile of (a) plane and (b) corroded mild steel substrates

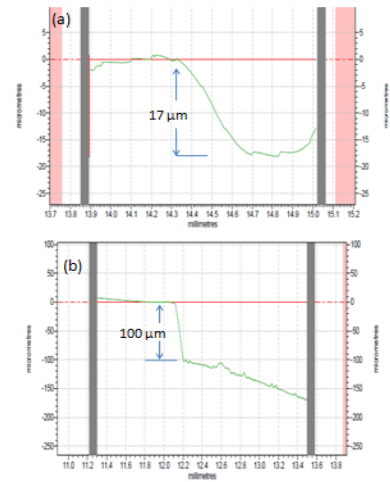

**Fig 14** Thickness of BMI/Al coatings depending upon time by surface profilometer

### Tensile tests

UTM tests were carried out in mild steel and aluminum samples. Typical stress-strain curves for MS and Al are shown in fig 15 & 16. The results obtained from the tensile tests are given in Table 1. In the case of mild steel the plastic deformation started at 349 MPa called the yield stress. It can be observed that the stress suddenly decreases from a point called the upper yield point while continued deformation fluctuates slightly about some constant stress value, termed the lower yield point. For metals that display this effect, the yield strength is taken as the average stress that is associated with the lower yield point. The ultimate tensile stress (UTS) was around 507 MPa at which the specimen failure occurred as shown in fig 15(a). The YS/UTS ratio is a parameter for strain hardening. The lower the ratio the higher will be the strain hardening. For MS samples the value was 0.69 while for Aluminum samples it was 0.979 which implies strain hardening being active in MS which matched with our previous results by Vickers microhardness tests.

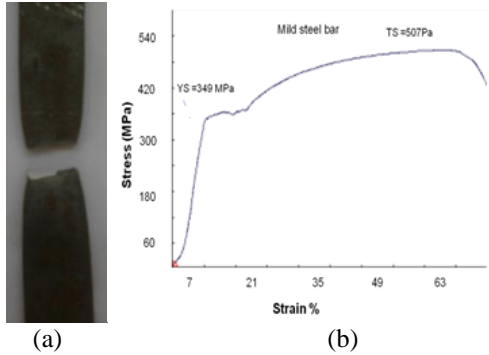

(a)             (b)

**Fig 15**: Stress-strain curve obtained from tensile tests done on Mild steel

In case of aluminum bar, the curve initially was parallel to the strain axis which was mainly due to slip in the grip. The failure associated with tensile stress was not at the gauge length but nearer to the grip due to flaw associated with sample design causing a stress concentration near the grip (Fig 16 (a))

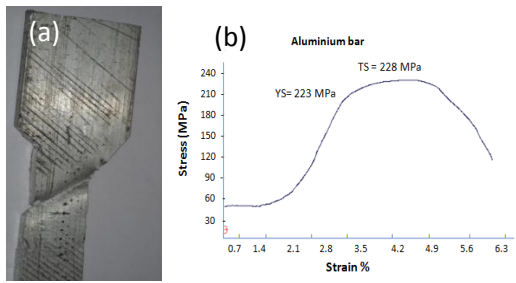

**Fig 16**: Stress-strain curve obtained from tensile tests done on (a) Aluminum bar (b) sample failure

**Table 1:** Tensile test results performed on Mild steel and (b) Aluminum bar

| Results | Mild steel bar | Aluminum bar |
|---|---|---|
| Maximum load ($F_{max}$) | 40.6 kN | 38.4 kN |
| Displacement at $F_{max}$ | 29.7 mm | 9.4 mm |
| Maximum displacement | 34.9 mm | 13.6 mm |
| Cross sectional area | 80.0 mm$^2$ | 168 mm2 |
| Tensile strength (TS) | 507 MPa | 228 MPa |
| % Elongation | 26% | 0.92% |
| Yield load | 27.9 kN | 37.6 kN |
| Yield stress | 349 MPa | 223 MPa |
| YS/UTS ratio | 0.69 | 0.979 |

**X-ray diffraction studies**

X-ray diffraction studies on BMI coatings on aluminum are shown in fig 17. The occurrence of sharp peaks is a clear conformation of crystalline nature of the films. The peaks were deconvoluted using Gaussian fit as shown in the fig 18. The fitting parameters are given in the table 2 below.

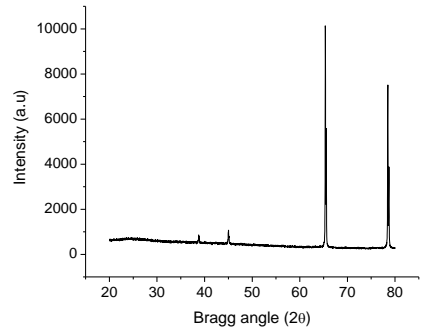

**Fig 17**: XRD plot of BMI coating on Al substrate

The crystallite size L can be determined using Scherrer equation. $L = k\lambda/\beta \cos\theta$, where k is the Scherer factor ~ 0.9. $\lambda$ is the X-ray wavelength which in this case is 1.54Å for Cu-$K\alpha$ radiation. 2θ is the Bragg angle and $\beta = (B^2 - b_o^2)^{1/2}$ is the pure line width. B is the experimental peak width and $b_o$ is the instrumental peak broadening factor which has a negligible value [13]. Hence for the sake of simplicity we are considering only the experimental peak broadening for determination of crystallite size. The crystallite sizes were found vary from 180 nm to 200 nm.

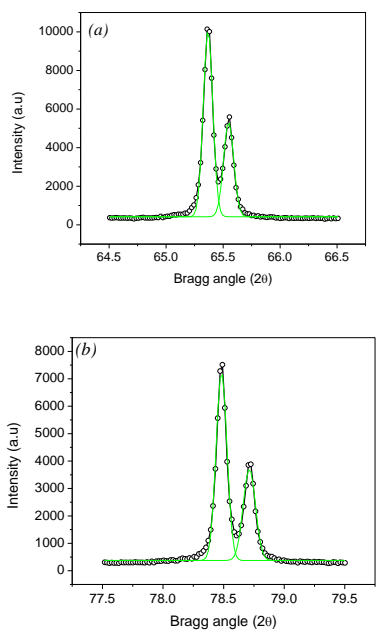

**Fig 18**: Deconvolution of peaks obtained from XRD at two different ranges

**Table 2: Parameter of the deconvolution of XRD peaks at different Bragg angles**

| Bragg angle (2θ) | Area | FWHM (B) | Intensity (a. u) | Crystallite size (nm) |
|---|---|---|---|---|
| 65.4 | 1004 | 0.083 | 9669 | 198 |
| 65.6 | 549 | 0.089 | 4906 | 185 |
| 78.5 | 771 | 0.089 | 6882 | 202 |
| 78.7 | 423 | 0.10 | 3340 | 180 |

**Contact angle measurements**

The contact angle measurements of pristine Al surface, plasma surface modified (PSM) Al surface and BMI coated Al surface are shown in fig 19 (a), (b) and (c) respectively. The reduction of contact angle in the PSM-Al surface compared to pristine Al surface indicates increase in hydrophilic nature of the surface which leads to better adhesion. The plasma surface modification was done by atmospheric pressure plasma jet (APPJ) details of which are published in ref [1]. BMI coated Al on the other hand showed hydrophobic nature or water repellent character evident from the increase contact angle.

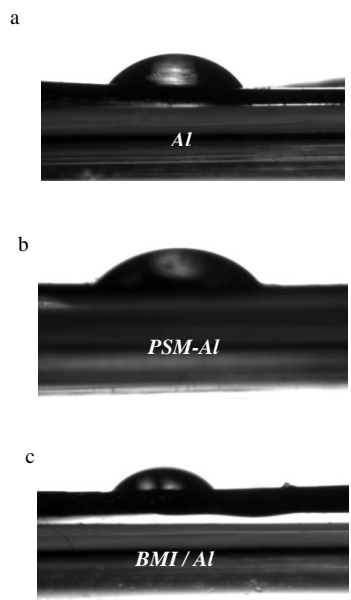

**Fig 19**: Contact angle measurements on the surface of (a) Al (b) PSM-Al and (c) BMI/Al

**Conclusions**

Bismaleimide (BMI) coatings deposited on aluminium (Al) and mild steel (MS) showed higher hardness compared to bare Al and MS. The coatings showed good corrosion resistance in Ringers solution. Metallization of BMI by Aluminum was done by vacuum thermal deposition to make Al/BMI/Al and Al/BMI/MS. Crystallinity was observed in the BMI coatings by XRD. Surface profilometer gave roughness and thickness of the coatings. Contact angle measurements showed hydrophobic nature of the coatings.

**Acknowledgments**
The authors thank Science and Engineering Research Board, India for research grant SERB/F/3482/2012-2013 (Dated 24 September 2012). The authors also thank Prof. S. Bhowmik for his suggestions and Dr. S.K.Mishra, CSIR-NML Jamshedpur for profilometer and hardness tests.